\documentclass[10pt]{iopart}
\usepackage{iopams}
\usepackage{setstack}
\usepackage{curves}
\usepackage{xcolor}
\usepackage{multirow}
\usepackage{graphicx,rotating}
\usepackage{collapse}

\makeatletter
\newcommand{\manuallabel}[2]{\def\@currentlabel{#2}\label{#1}}
\makeatother
\newcommand{\Exp}[1]{\mathrm{e}^{\mbox{\footnotesize$#1$}}}
\newcommand{\Eq}[2][equation~]{#1(\ref{eq:#2})}

\newcommand{\Ai}{{\rm{Ai}}}
\newcommand{\J}{{\rm{J}}}
\newcommand{\half}{\frac{1}{2}}
\newcommand{\I}{{\rm i}}
\newcommand{\ket}[1]{{\left|{#1}\right\rangle}}
\newcommand{\bra}[1]{{\left\langle{#1}\right|}}
\newcommand{\D}{{\rm d}}
\newcommand{\Dt}{\frac{\D}{\D t}}
\newcommand{\ds}{\displaystyle}
\newcommand{\Real}{\mathnormal{\rm{Re}}}
\newcommand{\pder}[2]{\frac{\partial #1}{\partial #2}}
\newcommand{\intr}{\int(\D\bi{r})\,}
\newcommand{\Int}[1][-5pt]{\int\limits_{\begin{picture}(16,3)(-8,-3)%
		\put(0,0){\curve(-3,0,-8,0)\curve(3,0,8,0)}%
		\put(8,0){\curve(0,0,-1.5,1.5)\curve(0,0,-1.5,-1.5)}%
		\put(0,0){\arc(-3,0){180}}\put(0,0){\makebox(0,0){$\cdot$}}%
		\end{picture}}\hspace*{#1}}
\newcommand{\grad}{\bnabla}

\newenvironment{eqnArray}[1][3em]{%
    \begin{equation}\fl\rule{#1}{0pt}\begin{array}[b]{rcl}}
   {\end{array}\end{equation}}

\def\numparts{\addtocounter{equation}{1}%
     \setcounter{eqnval}{\value{equation}}%
     \setcounter{equation}{0}%
     \def\theequation{\ifnumbysec
     \arabic{section}.\arabic{eqnval}{\alph{equation}}%
     \else\arabic{eqnval}{\alph{equation}}\fi}}
\def\endnumparts{\def\theequation{\ifnumbysec
     \arabic{section}.\arabic{equation}\else
     \arabic{equation}\fi}%
     \setcounter{equation}{\value{eqnval}}}

\begin{document}

\title[]{Systematic corrections to the Thomas--Fermi approximation %
	 without a gradient expansion}

\author{Thanh Tri Chau$^{1}$, Jun Hao Hue$^{1,\,2}$, %
        Martin-Isbj\"orn Trappe$^{1}$ and Berthold-Georg Englert$^{1,\,3,\,4}$}

\address{$^1$ Centre for Quantum Technologies, National University of %
	 Singapore, 3 Science Drive 2, Singapore 117543, Singapore}
\address{$^2$ Graduate School for Integrative Sciences \& Engineering, %
	 National University of Singapore, 28 Medical Drive, 
         Singapore 117456, Singapore}
\address{$^3$ Department of Physics, National University of Singapore, %
         2 Science Drive 3, Singapore 117542, Singapore}
\address{$^4$ MajuLab, CNRS-UNS-NUS-NTU International Joint Unit, %
         UMI 3654, Singapore}

\ead{cqtctt@nus.edu.sg{\normalfont,} %
     e0028667@u.nus.edu{\normalfont,} %
     martin.trappe@quantumlah.org {\normalfont and} cqtebg@nus.edu.sg}
\vspace{10pt}
\begin{indented}
\item[]Posted on the arXiv on 24 March 2018
\end{indented}

\begin{abstract}
We improve on the Thomas--Fermi approximation for the single-particle density
of fermions by introducing inhomogeneity corrections.
Rather than invoking a gradient expansion, we relate the density to the
unitary evolution operator for the given effective potential energy and
approximate this operator by a Suzuki--Trotter factorization.
This yields a hierarchy of approximations, one for each approximate
factorization.
For the purpose of a first benchmarking, we examine the approximate densities
for a few cases with known exact densities and observe a very satisfactory, and
encouraging, performance.
As a bonus, we also obtain a simple fourth-order leapfrog algorithm for the
symplectic integration of classical equations of motion.
\end{abstract}

\vspace{2pc}

\section{Introduction}
All practical applications of density-func\-tion\-al theory (DFT) to systems of
interacting particles require trustworthy approximations to the functionals
for the kinetic energy and the interaction energy or---more relevant for
the set of equations that one needs to solve self-consistently---to their
functional derivatives.
While the Kohn--Sham (KS) scheme \cite{Kohn+1:65} avoids approximations for the
kinetic-energy functional (KEF), this comes at the high price of a CPU-costly
solution of the eigenvalue-eigenstate problem for the effective
single-particle Hamiltonian.
The popular alternative to KS-DFT proceeds from the KEF in Thomas--Fermi (TF)
approximation \cite{Thomas:26,Fermi:27} and improves on that by the inclusion
of inhomogeneity corrections in the form of gradient terms, with the von
Weizs\"acker term \cite{Weizsacker:35} as the leading correction.

This gradient expansion is notorious for its lack of convergence (see, e.g.,
\cite{Sergeev+3:16a}), the wrong sign of the von Weizs\"acker term for
one-dimensional systems \cite{Holas+2:91}, and the vanishing of all
corrections for two-dimensional systems
\cite{Holas+2:91,Shao:93,Salasnich:07}---or so it seems \cite{Trappe+4:16}.
While cures have been suggested, such as the use of a Pad\'e approximant
rather than the power series (see, e.g., \cite{Sergeev+3:16b}), or the  partial
re-summation of the series with the aid of Airy-averaging techniques
\cite{Baltin:72,Durand+2:78,Englert+1:84,Englert:88,Trappe+3:17}, the
situation is hardly satisfactory. 

Recently, however, Ribeiro \textit{et al.}\ \cite{Ribeiro+4:15}
demonstrated that one can improve very substantially on the TF approximation
without any gradient expansion at all.
True, the method employed in \cite{Ribeiro+4:15} and related papers
\cite{Cangi+3:10,Ribeiro+1:17} is designed for one-dimensional problems and
has so far resisted all attempts of extending it to two and three-dimensional
situations. But this work strongly encourages the search for other approximation schemes
that do not rely on gradient expansions and are not subject to the said
limitations. 
We are here reporting one scheme of this kind.

Our approach is based on the reformulation of DFT in which the effective
single-particle potential energy $V(\bi{r})$ is a variational variable on equal
footing with the single-particle density $n(\bi{r})$
\cite{Englert:88,Englert:92}. 
All functionals of $V(\bi{r})$ can be stated in terms of the effective
single-particle Hamiltonian, and we relate them to the corresponding unitary
evolution operator.
That, in turn, is then systematically approximated by products of simpler
unitary operators of the Suzuki--Trotter (ST) kind---known as split-operator
approximations; see, e.g., \cite{Hatano+1:05}.
There is, in particular, a relatively simple five-factor approximation that is
correct to fourth order.
It promises a vast improvement over the TF approximation without the high
costs of the KS method and without the dimensional limitation of the Ribeiro
\textit{et al.}\ method. 

The unitary ST approximations directly provide symplectic approximants for
classical time evolution. The here developed higher-order ST approximations
therefore have the potential to significantly improve upon standard
second-order leapfrog algorithms \cite{Newton:87,Neal:11} or even fourth-order
Runge--Kutta methods.

\section{Single-particle density and evolution operator}
We consider a system of $N$ unpolarized spin-$\frac{1}{2}$ fermions of mass
$m$, subject to external forces that derive from the potential energy
$V_{\rm{ext}}(\bi{r})$. 
The energy functional
\begin{equation}\label{eq:1}
\fl\rule{3em}{0pt}  
    E[V,n,\mu]=E_1[V-\mu]-\intr[V(\bi{r})-V_{\mathrm{ext}}(\bi{r})]\,n(\bi{r})
               +E_{\mathrm{int}}[n]+\mu N\,,
\end{equation}
which has $n(\bi{r})$ and $V(\bi{r})$ as well as the chemical potential
$\mu$ as variables, is stationary at the actual values.
Therefore, we have the three variational equations
\numparts
\begin{eqnarray}
   \label{eq:2a}
   \makebox[5em][s]{${\delta V:}$ $n(\bi{r})$}
   &=&\frac{\delta}{\delta V(\bi{r})}E_1[V-\mu]\,,\\
   \label{eq:2b}
   \makebox[5em][s]{${\delta n:}$ $V(\bi{r})$}
   &=&V_{\mathrm{ext}}(\bi{r})+\frac{\delta}{\delta
        n(\bi{r})}E_{\mathrm{int}}[n]\,,\\
   \label{eq:2c}
   \makebox[5em][s]{${\delta \mu:}$ $N$}
   &=&-\frac{\partial}{\partial \mu}E_1[V-\mu]\,,    
\end{eqnarray}
\endnumparts
which we need to solve jointly.
Here, $E_{\mathrm{int}}[n]$ is the standard interaction-energy functional
(IEF) of the Hohen\-berg--Kohn (HK) theorem \cite{Hohenberg+1:64}, 
and the single-particle energy functional $E_1[V-\mu]$ is related to the
HK-KEF by a Legendre transformation,
\begin{equation}\label{eq:3}
E_1[V-\mu]=E_{\mathrm{kin}}[n]+ \intr[V(\bi{r})-\mu]\,n(\bi{r})\,.
\end{equation}
\Eq[Equation~]{2a} yields $n(\bi{r})$ for given $V(\bi{r})$ and $\mu$,
whereas we obtain $V(\bi{r})$ for given $n(\bi{r})$ from \Eq{2b};
and the correct normalization of $n(\bi{r})$ to the particle number $N$ is
ensured by \Eq{2c} because it implies
\begin{equation}
\label{eq:4}
\intr n(\bi{r})=N
\end{equation}
when combined with \Eq{2a}.

In the KS formalism, we have
\begin{equation}
\label{eq:5}
n(\bi{r})
=2\bra{\bi{r}}\eta{\bigl(\mu-H{\left(\bi{P},\bi{R}\right)}\bigr)}\ket{\bi{r}}
\end{equation}
with the eigenbras $\bra{\bi{r}}$ and eigenkets $\ket{\bi{r}}$ of the
position operator $\bi{R}$ and the single-particle Hamiltonian
\begin{equation}
\label{eq:6}
H(\bi{P},\bi{R})=\frac{1}{2m}\bi{P}^2+V(\bi{R})\,.
\end{equation}
Here and throughout the paper, the upper-case letters $\bi{P},\bi{R}$ denote
the quantum mechanical momentum and position operators, whereas the lower-case
letters $\bi{p},\bi{r}$ stand for their classical counterparts. 
We can either accept \Eq{5} as determining the right-hand side in \Eq{2a},
which is exact if we include the difference between the KEFs for interacting and
non-interacting particles in the IEF, or we regard \Eq{5} as stating an
approximation for the right-hand side in \Eq{2a}---an approximation that has
a very good track record.
Whichever point of view we adopt, we shall have to deal with \Eq{5}.

The task, then, is to evaluate, in a good approximation, the diagonal matrix
element of the step function of $\mu-H$ without computing the eigenvalues of
$H$ and the corresponding single-particle wave functions (``orbitals'').
In the tradition of the TF approximation and its refinements by gradient
terms, and also in the spirit of the Ribeiro \textit{et al.}\ work, we target
an orbital-free (OF) formalism. 
In a first step toward this goal, we relate $n(\bi{r})$ to the unitary
evolution operator 
$\exp{\left(-\frac{\I t}{\hbar} H{\left(\bi{P},\bi{R}\right)}\right)}$,
\begin{equation}\label{eq:7}
   n(\bi{r})=2\Int\frac{\D t}{2\pi\I t}\,
             \exp{\left(\frac{\I t}{\hbar}\mu\right)}\,
             \bra{\bi{r}}
             \exp{\left(-\frac{\I t}{\hbar}H(\bi{P},\bi{R})\right)}
             \ket{\bi{r}}\,,
\end{equation}
where the integration path from $t=-\infty$ to $t=\infty$ crosses the
imaginary $t$ axis in the lower half-plane
\cite{Golden:57,Light+1:73,Lee+1:75}. 
Therefore, an approximation for 
$\exp{\left(-\frac{\I t}{\hbar} H{\left(\bi{P},\bi{R}\right)}\right)}$
provides a corresponding approximation for $n(\bi{r})$.

\section{Hierarchy of ST approximations---TF and beyond}
\begin{table}
\caption{\label{tab:1}%
  Summary of the hierarchy of systematic ST approximations of the
  time-evolution operator discussed in sections
  \ref{sec:ST3}--\ref{sec:STdensities}. 
  We require these approximations to be exact for the constant-force
  potential, and be at least of third order in time $t$ for consistency with
  the leading gradient correction. 
  The fourth-order approximation $U_7\mbox{\large$|$}_{\epsilon\to 0}$ derived in
  section \ref{sec:U7eps} meets all these requirements.} 
\begin{indented}
\item[]\begin{tabular}{@{}lllllll}
\br
&$U_3$&$U_{3'}$&$U_5$&$U_5\bigr|_{z\to\infty}$&$U_7$&$U_7\bigr|_{\epsilon\to0}$\\
\mr
\multirow{2}{3.5cm}{Exact for constant-force potential}&\multirow{2}{*}{no}
&\multirow{2}{*}{no}&\multirow{2}{*}{\textbf{yes}}
&\multirow{2}{*}{\textbf{yes}}&\multirow{2}{*}{\textbf{yes}}
&\multirow{2}{*}{\textbf{yes}}\\
&&&&&&\\
\mr
\multirow{2}{3.5cm}{Correct up to order}&\multirow{2}{*}{2nd}
&\multirow{2}{*}{2nd}&\multirow{2}{*}{2nd}&\multirow{2}{*}{2nd}
&\multirow{2}{*}{2nd}&\multirow{2}{*}{\textbf{4th}}\\
&&&&&&\\
\mr
\multirow{2}{3.5cm}{Consistent with leading gradient correction}
&\multirow{2}{*}{no}&\multirow{2}{*}{no}&\multirow{2}{*}{no}
&\multirow{2}{*}{no}&\multirow{2}{*}{no}&\multirow{2}{*}{\textbf{yes}}\\
&&&&&&\\
\br
\end{tabular}
\end{indented}
\end{table}
\noindent In sections \ref{sec:ST3}--\ref{sec:STdensities}, we derive explicit
approximations of the time-evolution operator and the according densities
based on \Eq{7}. 
These approximations successively incorporate exact constraints, eventually
leading to a five-term factorization $U_7\bigr|_{\epsilon\to 0}$ that captures
the exact evolution operator up to the fourth order in time $t$, is exact for
the constant-force potential, and is consistent with the leading gradient
correction. 
Table \ref{tab:1} summarizes our results.

\subsection{Recovering the TF approximation---ST3}\label{sec:ST3}
For a $D$-di\-men\-sion\-al system ($D=1$, $2$, or $3$), we have the TF
approximation
\begin{equation}
\label{eq:8}
n^{\ }_{\rm{TF}}(\bi{r})= 2\int\!\frac{(\D\bi{p})}{(2\pi\hbar)^D}\,
\eta{\bigl(\mu-H{\left(\bi{p},\bi{r}\right)}\bigr)}\,.
\end{equation}
As one verifies easily, this is recovered by the three-factor ST approximation
(ST3) 
\begin{eqnarray}
\label{eq:9}
\fl\rule{0.3em}{0pt}
\exp{\left(-\frac{\I t}{\hbar} H{\left(\bi{P},\bi{R}\right)}\right)}
&=&\exp{\left(-\frac{\I t}{\hbar}
        {\left[\frac{1}{2m}\bi{P}^2+V(\bi{R})\right]}\right)}
\\\nonumber&\simeq&\exp{\left(-\frac{\I t}{2\hbar}V(\bi{R})\right)}
\exp{\left(-\frac{\I t}{2\hbar m}\bi{P}^2\right)}
\exp{\left(-\frac{\I t}{2\hbar}V(\bi{R})\right)}\equiv U_3\,,
\end{eqnarray}
for which
\begin{equation}
\label{eq:10}
\bra{\bi{r}}U_3\ket{\bi{r}}
=\int\!\frac{(\D\bi{p})}{(2\pi\hbar)^D}\,
\exp{\left(-\frac{\I t}{\hbar}
     {\left[\frac{1}{2m}\bi{p}^2+V(\bi{r})\right]}\right)}\,.
\end{equation}

We note that the symmetrized approximation of \Eq{9} is not needed here,
the two-factor approximations 
$\exp{\left(-\frac{\I t}{\hbar}V(\bi{R})\right)}%
\exp{\left(-\frac{\I t}{\hbar}\frac{1}{2m}\bi{P}^2\right)}$ or
$\exp{\left(-\frac{\I t}{\hbar}\frac{1}{2m}\bi{P}^2\right)}%
\exp{\left(-\frac{\I t}{\hbar}V(\bi{R})\right)}$ would work just as well, and
so would other asymmetric ways of sandwiching the kinetic-energy factor by
potential-energy factors.
Yet, the symmetric version does have an advantage: An approximation $U(t)$,
\begin{equation}
\label{eq:11}
\exp{\left(-\frac{\I t}{\hbar}{\left[\frac{1}{2m}\bi{P}^2
                                     +V(\bi{R})\right]}\right)}
\simeq U(t)\,,
\end{equation}
with the property $U(t)U(-t)=1$, as is the case for $U_3$, 
cannot have a leading error proportional to an even power of $t$.
Since ST3 is obviously correct to first order in $t$, the terms $\propto t^2$
must also be correct; indeed, the leading error is of order $t^3$. 
In this sense, ST3 is a second-order approximation.

If we interchange the roles of the kinetic and the potential energy in \Eq{9},
we obtain another three-factor approximation of second order (ST3'),
\begin{equation}
\label{eq:9'}
U_{3'}=\exp{\left(-\frac{\I t}{4\hbar m}\bi{P}^2\right)}
\exp{\left(-\frac{\I t}{\hbar}V(\bi{R})\right)}
\exp{\left(-\frac{\I t}{4\hbar m}\bi{P}^2\right)}\,.
\end{equation}
This gives
\begin{equation}
\label{eq:n3p}
\fl\rule{3.5em}{0pt} 
n_{3'}(\bi{r})=2\int\!\frac{(\D\bi{p}_1)(\D\bi{p}_2)(\D\bi{r}_1)}
{(2\pi\hbar)^{2D}}\,
\exp{\left(\frac{\I}{\hbar}\bi{r}_1\cdot(\bi{p}_1-\bi{p}_2)\right)}
\eta(\mu-\mathcal{H}_{3'})
\end{equation}
with
\begin{equation}
\label{eq:8''}
\mathcal{H}_{3'}=\frac{\bi{p}_1^2+\bi{p}_2^2}{4m}+V(\bi{r}+\bi{r}_1)
\end{equation}
for the corresponding approximation for $n(\bi{r})$.

\subsection{Beyond the TF approximation---ST5}
Since $U_3$ and $U_{3'}$ of \Eq[equations~]{9} and \Eq[]{9'}
have errors for a constant-force situation,
$ \grad V(\bi{r})=\rm{constant}$, they do not incorporate Langer's correction
\cite{Langer:37} which improves the description enormously at the border
between the classically allowed and forbidden regions.
Langer's correction is a key ingredient in the Ribeiro \textit{et al.}\ 
method \cite{Ribeiro+4:15}.
With this in mind we now consider a five-factor approximation (ST5) of the form
\begin{eqnArray}\label{eq:12}
U_5&=&\ds\exp{\left(-\frac{\I t}{2\hbar m}y_1\bi{P}^2\right)}
       \exp{\left(-\frac{\I t}{\hbar}x_1V(\bi{R})\right)}
       \exp{\left(-\frac{\I t}{2\hbar m}y_0\bi{P}^2\right)}\\[2ex]
&&\ds\times\exp{\left(-\frac{\I t}{\hbar}x_2V(\bi{R})\right)}
         \exp{\left(-\frac{\I t}{2\hbar m}y_2\bi{P}^2\right)}
\end{eqnArray}
and require that it is exact for vanishing dyadic $\grad\grad V(\bi{r})$.
This identifies a set of admissible coefficients,
\begin{equation}\label{eq:13}\fl\rule{5em}{0pt}
\left.\begin{array}{c} x_1 \\ x_2 \end{array}\right\}
=\frac{1}{2}(1\pm z)\,,\qquad 
y_0=\frac{2/3}{1-z^2}\,,\qquad
\left.\begin{array}{c} y_1 \\ y_2 \end{array}\right\}
=\frac{1\pm3z}{6(1\pm z)}\,,
\end{equation}
parameterized by $z$. 
While $U_3$ and $U_{3'}$ are particular cases of $U_5$, they are not in the
$z$-parameterized set.

The error in $U_5$ is of third order; the leading error term is the double
commutator $\Delta_5\frac{(-\I t/\hbar)^3}{(2m)^2}%
\Bigl[\bi{P}^2,\bigl[\bi{P}^2,V(\bi{R})\bigr]\Bigr]$ 
with
\begin{equation}
\label{eq:17}
\Delta_5=\frac{1}{72}\frac{1+3z^2}{1-z^2}
=\cases{\hphantom{+}\frac{1}{72}&for $z=0$\,,\\
\hphantom{+}\frac{1}{48}&for $z=\pm\case{1}{3}$\,,\\
-\frac{1}{24}&for $z\to\infty$\,.\\}
\end{equation}
Since $\Delta_5\neq0$ for all real $z$ values ($z=\I/\sqrt{3}$ is not an option;
see below), we need weighted sums such as
\begin{equation}
\label{eq:18}
\fl\rule{4em}{0pt} 3\,U_5(z=0)-2\,U_5(z=\case{1}{3})\quad{\mathrm{or}}\quad 
\frac{3}{4}\,U_5(z=0)+\frac{1}{4}\,U_5(z=\infty)
\end{equation}
for a third-order approximation, but then the unitary evolution operator is
approximated by the non-unitary sum of two unitary operators, and this should
better be avoided \cite{Hatano+1:05}.

The resulting approximation for $n(\bi{r})$ is
\begin{eqnArray}[5em]\label{eq:n5}
 n_5(\bi{r})&=&\ds 2
\int\frac{(\D\bi{p}_0)(\D\bi{p}_1)(\D\bi{p}_2)(\D\bi{r}_1)(\D\bi{r}_2)}
{(2\pi\hbar)^{3D}}
\,\eta(\mu-\mathcal{H}_5)\\[3ex]
&&\ds\hphantom{2\int}
\times\exp{\left(\frac{\I}{\hbar}{\bigl[\bi{r}_1\cdot(\bi{p}_0-\bi{p}_1)
	+\bi{r}_2\cdot(\bi{p}_2-\bi{p}_0)\bigr]}\right)}
\end{eqnArray}
where
\begin{equation}
\label{eq:20}
\mathcal{H}_5
=\frac{y_0\bi{p}_0^2+y_1\bi{p}_1^2+y_2\bi{p}_2^2}{2m}
+x_1V(\bi{r}+\bi{r}_1)+x_2V(\bi{r}+\bi{r}_2)
\end{equation}
is the sum of a kinetic-energy term and a potential-energy term.
For this kinetic energy to be positive, we need $y_0,y_1,y_2\geq0$,
hence $-\frac{1}{3}\leq z\leq\frac{1}{3}$.

Although the $z$ values are outside of this range in the ${z\to\infty}$ limit
in \Eq{17}, this does provide an acceptable approximation,
namely
\begin{eqnArray}[4.5em]
\label{eq:U5inf}
\ds U_5\Bigr|_{z\to\infty}
&=&\ds\exp{\left(-\frac{\I t}{4\hbar m}\bi{P}^2\right)}
\exp{\left(-\frac{\I t}{\hbar}\left(V
	-\frac{1}{12m}{\left[t\grad V\right]}^2\right)(\bi{R})\right)}
\\[2ex]
&&\ds\times\exp{\left(-\frac{\I t}{4\hbar m}\bi{P}^2\right)}\,,
\end{eqnArray}
which is a three-factor approximation for the evolution operator.
The corresponding expression for the density is
\begin{eqnArray}
\label{eq:22}
n_5(\bi{r})\Bigr|_{z\to\infty}
&=&\ds 2\int\D x\,\Ai(x)
\int\frac{(\D\bi{p}_1)(\D\bi{p}_2)(\D\bi{r}_1)}{(2\pi\hbar)^{2D}} 
\,\eta{\left(\mu-\mathcal{H}_5^{(x)}\right)}\\[3ex]
&&\ds\hphantom{2\int\D x\,\Ai(x)\ \ \ }
  \times\exp{\left(\frac{\I}{\hbar}(\bi{r}-\bi{r}_1)
                      \cdot(\bi{p}_1-\bi{p}_2)\right)}
\end{eqnArray}
with the Airy function $\Ai(\ )$ and
\begin{equation}
\label{eq:23}
\mathcal{H}_5^{(x)}=\frac{\bi{p}_1^2+\bi{p}_2^2}{4m}+V(\bi{r}_1)
+\frac{x}{2}{\left(\frac{2}{m}{\bigl[\hbar\grad V(\bi{r}_1)\bigr]}^2
\right)}^{1/3}\,.
\end{equation}
We emphasize that there is \emph{no gradient expansion} in \Eq{U5inf}; 
the appearance of $\grad V$ is a consequence of
\begin{eqnArray}[2em]
\label{eq:24}
&&\ds\exp{\left(-\frac{\I t z}{2\hbar}V(\bi{R})\right)}
\exp{\left(\frac{\I t}{3\hbar mz^2}\bi{P}^2\right)}
\exp{\left(\frac{\I t z}{2\hbar}V(\bi{R})\right)}\\[3ex]
&=&\ds\exp{\left(\frac{\I t}{3\hbar mz^2}
           {\left[\bi{P}+\half tz\grad V(\bi{R})\right]}^2\right)}
\mathrel{\begin{array}[b]{c}{}^{z\to\infty}\\[-2ex]
	-\!\!\!-\!\!\!-\!\!\!\longrightarrow\end{array}}
\exp{\left(\frac{\I t}{12\hbar m}\bigl[t\grad V(\bi{R})\bigr]^2\right)}\,.
\end{eqnArray}

\subsection{Beyond the TF approximation---ST7}
Since $U_5$ has a nonzero third-order error coefficient, ${\Delta_5\neq0}$, it
does not account in full for the leading gradient correction, unless we resort
to the non-unitary approximations of \Eq{18}.
ST5 also fails to reproduce the leading correction to the Wigner function of
the evolution operator (see, e.g., \cite{Trappe+3:17}).
Let us, therefore, consider a seven-factor approximation (ST7) of the form
\begin{equation}
\label{eq:25}
U_7=\exp{\left(-\frac{\I t}{\hbar}x_3V(\bi{R})\right)} U_5
\exp{\left(-\frac{\I t}{\hbar}x_4V(\bi{R})\right)}
\end{equation}
for which
\begin{eqnArray}\label{eq:26}
n_7(\bi{r})&=&\ds 2
\int\frac{(\D\bi{p}_0)(\D\bi{p}_1)(\D\bi{p}_2)(\D\bi{r}_1)(\D\bi{r}_2)}
{(2\pi\hbar)^{3D}}\,\eta(\mu-\mathcal{H}_7)\\[3ex]
&&\ds\hphantom{2\int}
\times\exp{\left(\frac{\I}{\hbar}{\bigl[\bi{r}_1\cdot(\bi{p}_0-\bi{p}_1)
	+\bi{r}_2\cdot(\bi{p}_2-\bi{p}_0)\bigr]}\right)}
\end{eqnArray}
with
\begin{equation}
\label{eq:27}
\mathcal{H}_7=\mathcal{H}_5+(x_3+x_4)V(\bi{r})\,.
\end{equation}
Here, we need ${x_1+x_2+x_3+x_4=1}$ and
\begin{equation}
y_0=\frac{3z_1z_2+1}{3(z_0-z_1)(z_0-z_2)}
\end{equation}
and the cyclic analogs for $y_1$ and $y_2$ with
\begin{equation}
\label{eq:29}\fl\rule{5em}{0pt}
z_0=x_1-x_2+x_3-x_4\,,\ \quad
z_1=2x_3-1\,,\ \quad
z_2=1-2x_4\,.
\end{equation}
\noindent to ensure that Langer's correction is incorporated.
The third-order error coefficient is now
\begin{equation}
\label{eq:30}
\Delta_7=\frac{1}{12}-\half x_1 y_1(1-y_1)-\half x_2 y_2(1-y_2)\,.
\end{equation}
We return to ST5 for $x_3=x_4=0$, $x_1-x_2=z$.

When insisting on $y_0,y_1,y_2\geq0$ to ensure positive kinetic-energy terms
in $\mathcal{H}_7$, the minimal value of $\Delta_7$ is 
positive, $\Delta_7\geq\frac{1}{12}-\frac{1}{18}\sqrt{2}=0.0048$.
While this lower bound is below that of $\Delta_5$,
$\Delta_5\geq\frac{1}{72}=0.0139$, the improvement is quantitative, not
qualitative. However, in the following section we demonstrate that
$\Delta_7=0$ in the limit $y_0\to0^-$, of the kind in \Eq{U5inf}.

\subsection{A fourth-order approximation}\label{sec:U7eps}
The symmetric version of this limit uses
\begin{equation}
\label{eq:31}
\left.\begin{array}{c} x_1\\x_2\end{array}\right\}
=\frac{1}{3}\pm\frac{1}{\epsilon}\,,\ \quad x_3=x_4=\frac{1}{6}
\end{equation}
for which
\begin{equation}
\label{eq:32}
y_0=-\frac{\epsilon^2}{36}+O(\epsilon^4)\,,\qquad 
\left.\begin{array}{c} y_1\\y_2\end{array}\right\}
=\frac{1}{2}\mp\frac{\epsilon}{24}+O(\epsilon^2)
\end{equation}
and $\Delta_7=-\epsilon^2/(12)^3+O(\epsilon^4)$.
For $\epsilon\to0$, then, the error coefficient vanishes, $\Delta_7\to0$, and
we obtain 
\begin{eqnArray}[6em]
\label{eq:U7inf}
U_7\Bigr|_{\epsilon\to0}&=&\ds\exp{\left(-\frac{\I t}{6\hbar}V(\bi{R})\right)}
\exp{\left(-\frac{\I t}{4\hbar m}\bi{P}^2\right)}\\[3ex]&&\ds\times
\exp{\left(-\frac{\I t}{\hbar}{\left(\frac{2}{3}V
	-\frac{1}{72m}[t\grad V]^2\right)}(\bi{R})\right)}\\[3ex]
&&\ds\times\exp{\left(-\frac{\I t}{4\hbar m}\bi{P}^2\right)}
\exp{\left(-\frac{\I t}{6\hbar}V(\bi{R})\right)}
\end{eqnArray}
as well as
\begin{eqnArray}[2em]
\label{eq:n7inf}
 n_7(\bi{r})\Bigr|_{\epsilon\to0}&=&\ds2\int\D x\,\Ai(x)
\int\frac{(\D\bi{p}_1)(\D\bi{p}_2)(\D\bi{r}_1)}{(2\pi\hbar)^{2D}}\,
\exp{\left(\frac{\I}{\hbar}(\bi{r}-\bi{r}_1)\cdot(\bi{p}_1-\bi{p}_2)\right)}
\\[3ex]&&\ds\rule{0em}{0pt}\times    
\eta{\left(\mu-\mathcal{H}_7^{(x)}\right)}\\
\end{eqnArray}
where 
\begin{equation}
\label{eq:35}\fl\rule{4em}{0pt}
\mathcal{H}_7^{(x)}=\frac{\bi{p}_1^2+\bi{p}_2^2}{4m}+\frac{1}{3}V(\bi{r})
+\frac{2}{3}V(\bi{r}_1)
+\frac{x}{2}{\left(\frac{1}{3m}\bigl[\hbar\grad V(\bi{r}_1)\bigr]^2
\right)}^{1/3}\,.
\end{equation}
The right-hand side in \Eq{U7inf} has the symmetry property discussed at \Eq{11}
and, as a consequence, is a fourth-order approximation to the evolution
operator. It also reproduces the leading correction to the Wigner function of
$U(t)$; see appendix \ref{Wigner}. 
The simplicity of the expression is striking if one compares it with the
fourth-order approximations in equations (44) and (144) in \cite{Hatano+1:05};
both have eleven factors.

\subsection{Approximate densities}\label{sec:STdensities}
In this section, we present numerically tractable expressions for the
approximate densities in equations \Eq[]{n3p}, \Eq[]{n5}, \Eq[]{22},
\Eq[]{26}, and \Eq[]{n7inf}. 
Upon performing the $\bi{p}$ integrations in \Eq{n3p}, we have
\begin{equation}\label{eq:36}
n_{3'}(\bi{r})=2\int(\D\bi{a}){\left(\frac{k_{3'}}{2\pi a}\right)}^D
\J_D{\left(2ak_{3'}\right)}
\end{equation}
where $\J_{D}(\ )$ is the Bessel function of order $D$ and 
\begin{equation}
\label{eq:37}
k_{3'}=\frac{1}{\hbar}{\Bigl[2m\mu-2mV(\bi{r}+\bi{a})\Bigr]}^{1/2}_+
\end{equation}
is the effective Fermi wave number, 
with ${\left[x\right]}^{1/2}_+=\eta(x)\sqrt{x}$.
Similarly, the single-particle density obtained from $U_5$ is
\begin{equation}
\label{eq:38}
n_5(\bi{r})=2\int\frac{(\D\bi{r}_1)(\D\bi{r}_2)}{(y_0y_1y_2)^{\frac{1}{2}D}}
{\left(\frac{k_{5}}{2\pi a}\right)}^{\frac{3}{2}D}
\J_{\frac{3}{2}D}{\left(ak_{5}\right)}
\end{equation}
where
\begin{equation}
\label{eq:39}
a^2=\frac{1}{y_0}(\bi{r}_1-\bi{r}_2)^2
+\frac{1}{y_1}(\bi{r}-\bi{r}_1)^2
+\frac{1}{y_2}(\bi{r}-\bi{r}_2)^2
\end{equation}
and
\begin{equation}
\label{eq:40}
k_5=\frac{1}{\hbar}
{\Bigl[2m{\bigl(\mu-x_1V(\bi{r}_1)-x_2V(\bi{r}_2)\bigr)}\Bigr]}^{1/2}_+\,.
\end{equation}
The right-hand side of \Eq{38} applies also for $n_7(\bi{r})$ with $k_5$
replaced by
\begin{equation}
\label{eq:41}
k_7=\frac{1}{\hbar}
{\Bigl[2m{\bigl(\mu-x_1V(\bi{r}_1)-x_2V(\bi{r}_2)-(x_3+x_4)V(\bi{r})\bigr)}
\Bigr]}^{1/2}_+.
\end{equation}
The density for the fourth-order approximation is
\begin{equation}
\label{eq:42}
n_7(\bi{r})\Bigr|_{\epsilon\to0}
=2\int\D x\,\Ai(x)\int(\D\bi{a})\,
{\left(\frac{k^{(x)}_{7}}{2\pi a}\right)}^D
\J_D{\left(2ak^{(x)}_{7}\right)}
\end{equation}
with
\begin{equation}
\label{eq:43}
\fl\rule{1em}{0pt} 
k^{(x)}_{7}=\frac{1}{\hbar}{\left[2m{\left(\mu-\frac{1}{3}V(\bi{r})
-\frac{2}{3}V(\bi{r}+\bi{a})\right)}
-x{\left(\frac{1}{3}{\bigl[\hbar m\grad V(\bi{r}+\bi{a})\bigr]}^2
\right)}^{1/3}\right]}^{1/2}_+;
\end{equation}
see appendix \ref{nDerivations} for detailed derivations of $n_7$ and 
$n_7\bigr|_{\epsilon\to0}$. 
While the right-hand side of \Eq{42} makes contact with equations \Eq[]{36}
and \Eq[]{38}, an equivalent expression for more efficient numerical
implementation is derived in appendices \ref{nDerivations} and
\ref{KDnumerics}. 
When we replace $k^{(x)}_{7}$ on the right-hand side of \Eq{42} by
\begin{equation}
\label{eq:44}\fl\rule{5em}{0pt}
k^{(x)}_{5}=\frac{1}{\hbar}{\left[2m{\bigl(\mu-V(\bi{r}+\bi{a})\bigr)}
-x{\left(2{\bigl[\hbar m\grad V(\bi{r}+\bi{a})\bigr]}^2
\right)}^{1/3}\right]}^{1/2}_+,
\end{equation}
we obtain the $z\to\infty$ limit of $n_5(\bi{r})$. 

We recover the TF density in \Eq{8} upon replacing $k_{3'}$, $k_{5}$,
$k_{5}^{(x)}$, $k_{7}$, or $k_{7}^{(x)}$ by $\ds k_{\mathrm{TF}}^{\ }=%
\frac{1}{\hbar}{\Bigl[2m{\bigl(\mu-V(\bi{r})\bigr)}\Bigr]}^{1/2}_+$,
respectively. 
All these refined approximations for $n(\bi{r})$ are ``nonlocal'' in the sense
that they involve an integration over a vicinity of the reference position
$\bi{r}$. 
They share this feature with the approximation introduced by Ribeiro
\textit{et al.}\ in \cite{Ribeiro+4:15} but, right now, it is quite
unclear how the two approaches are related to each other and whether one of the
approximations can be derived from the other.

\begin{figure}
\begin{center}
\includegraphics[width=\linewidth]{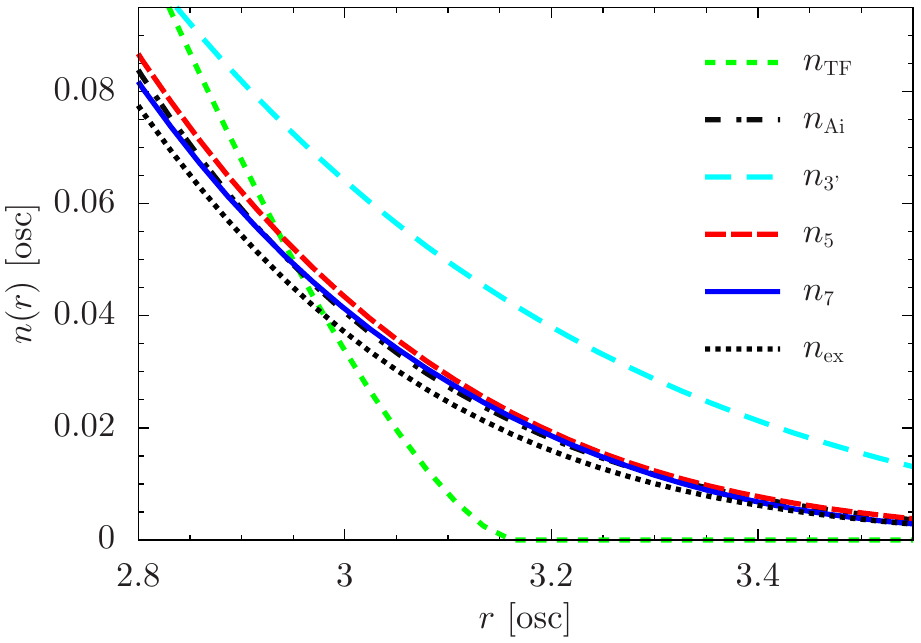}
\caption{\label{ClassQuantBorder}%
   Comparison of particle densities in Suzuki-Trotter approximations 
   in three-term, five-term, and seven-term factorizations ($n_{3'}$,
   $n_5$, and $n_7\bigr|_{\epsilon\to0}$)
   with the   Thomas--Fermi and the exact density ($n_{\rm{TF}}$ and
   $n_{\rm{ex}}$) as well as the approximation $n_{\rm{Ai}}$ obtained
   from a gradient expansion with Airy averaging. 
   We employ the potential energy ${V(\bi{r})=\half m\omega^2\bi{r}^2}$ for
   fermions in a three-dimensional isotropic harmonic oscillator, with
   $\mu=5\hbar\omega$ resulting in about 41 particles. 
   The ST approximations yield significant improvements over $n_{\rm{TF}}$,
   illustrated here for the region around the quantum-classical boundary along
   the radial coordinate $r$. 
   Harmonic oscillator units [osc] are used (\textsl{vulgo} 
   ${\hbar=m=\omega=1}$).}
\end{center}
\end{figure}

\section{Examples}
\begin{figure}
\begin{center}
\includegraphics[width=\linewidth]{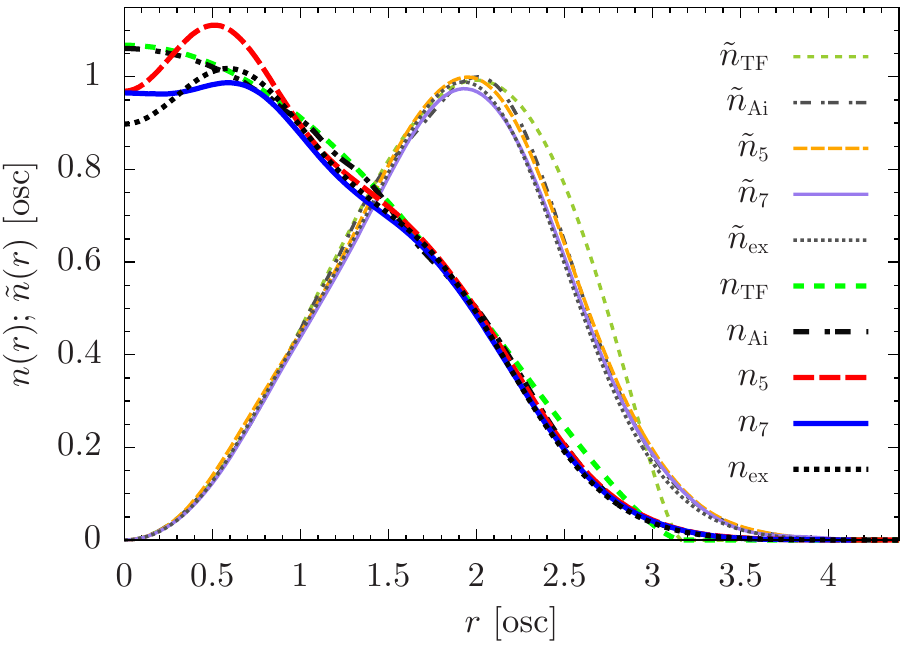}
\caption{\label{AllAndScaled}%
   Overview of particle densities with the same parameters as in
   figure~\ref{ClassQuantBorder}, together with the scaled densities 
   $\tilde{n}(r)=\half r^2n(r)$ that emphasize spatial regions of importance
   for integrated observables.} 
\end{center}
\end{figure}

In the following we assess the quality of the approximate densities presented
in section \ref{sec:STdensities}. 
Figure~\ref{ClassQuantBorder} illustrates the systematic improvement of the
ST approximation in \Eq{n7inf} over the TF approximation in \Eq{8}.
The plot also shows the densities associated with the three-factor and
five-factor ST approximations $n_{3'}$ and $n_5$; see \Eq{n3p} and \Eq{n5}
with ${z=0}$. 
The isotropic single-particle densities $n(r)$ of spin-$\half$ fermions with
harmonic potential energy in ${D=3}$ dimensions are obtained via
\Eq[equations~]{2a} and \Eq[]{2b} for a fixed chemical potential
${\mu=5\hbar\omega}$. 
In contrast to the quasi-classical $n_{\mathrm{TF}}$, $n_{3'}$ decays smoothly
across the boundary between classically allowed and forbidden regions, and the
higher-order ST approximations approach the exact density $n_{\mathrm{ex}}$
with successively higher accuracy. 
The results for $n_7$ are in line with the Airy-averaged density
$n_{\mathrm{Ai}}$ that also captures the leading quantum corrections, albeit
via a systematic gradient expansion (the densities $n_{\mathrm{Ai}}$ in
figures \ref{ClassQuantBorder} and \ref{AllAndScaled} are calculated for a
tempurature of one pico-Kelvin, i.e., very close to the ground state); see
\cite{Trappe+3:17}. 
$n_{5}$ and $n_{7}$ capture the quantum oscillations very well away from the
trap center (see figure~\ref{AllAndScaled}), while their deviations from the
exact density at small radii are less essential for global quantities like
energy or particle number; cf.~the weighted densities 
$\tilde{n}(r)=\half r^2n(r)$ in figure~\ref{AllAndScaled}.

\begin{figure}
\begin{center}
\includegraphics[width=\linewidth]{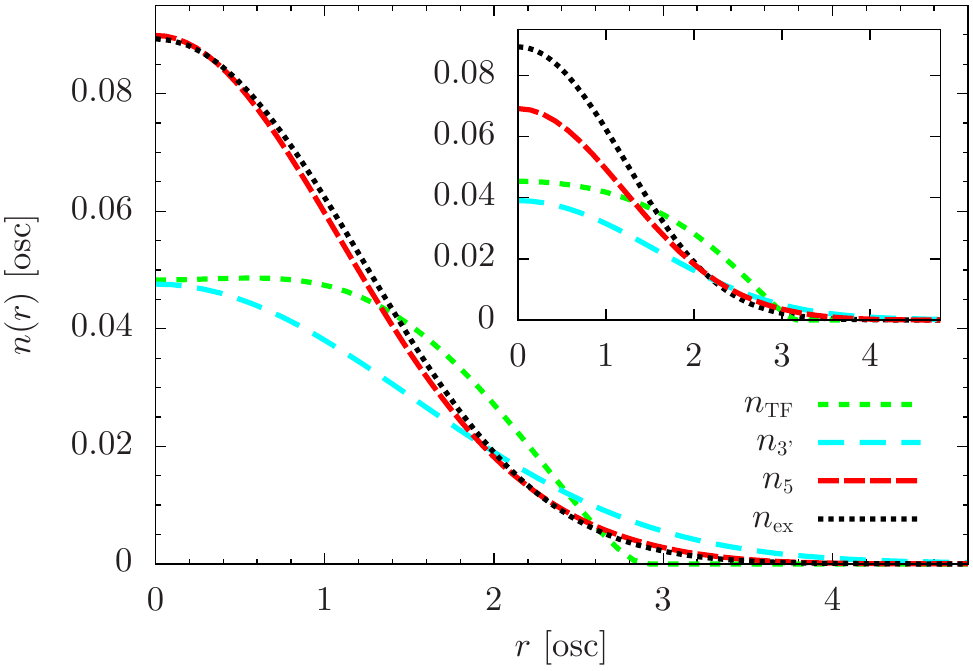}
\caption{\label{HookeAtom}%
   Isotropic particle densities for the Hooke atom. 
   Main figure: ST densities $n_{\mathrm{TF}}$, $n_{3'}$, and $n_{5}$, with the
   functional derivative of the interaction energy in \Eq{2b} evaluated at the
   exact density. 
   The excellent performance of $n_{5}$ is evident. 
   Inset: Converged densities from the self-consistently solved
   \Eq[equations~]{2a}--\Eq[]{2c}, incorporating the quasi-classical Hartree
   interaction energy $E^{\mathrm{(H)}}_{\mathrm{int}}[n]$ whose approximate
   nature is responsible for the differences to the main plot. 
We set $\hbar=m=2\omega=1$.}
\end{center}
\end{figure}

A fair comparison of the performance of the various ST approximations in the
context of \Eq[equations~]{2a}--\Eq[]{2c} requires fixed $\mu$ and $V$, although
this yields (slightly) different particle numbers for $n_{3'}$, $n_{5}$, and
$n_{7}$; 
from the physical point of view, $\mu$ is an auxilliary quantity
whose sole purpose is to fix the particle number $N$ for an actual physical
system. 
In figure~\ref{HookeAtom}, we find excellent agreement between $n_{5}$ and the
exact density for the Hooke atom of two Coulomb-interacting electrons in
external harmonic confinement \cite{Kestner+1:62,Kais+4:93}, although this
semiclassical approximation should not be expected to excel for systems with
such small particle numbers. We shall also consider $n_7$ for the Hooke atom,
among other systems, as soon as the numerics is optimized. 
When approximating the interaction contribution to $V(\bi{r})$ by the Hartree
term 
\begin{equation}
\label{eq:EintH}
\frac{\delta}{\delta n(\bi{r})}E^{\mathrm{(H)}}_{\mathrm{int}}[n]
\propto\int(\D\bi{r}')
\frac{n(\bi{r}')}{\boldsymbol{|}\bi{r}-\bi{r}'\boldsymbol{|}}\,,
\end{equation}
which is void of exchange and correlation contributions, we obtain the
densities in the inset of figure~\ref{HookeAtom} via the self-consistent
solution of \Eq[equations~]{2a}--\Eq[]{2c}. 
The disparity in results between the inset and the main figure serves as a
reminder that a truly systematic description requires expressions for particle
density and interaction functional at the same level of approximation.

\section{Interaction energy functional}
The various approximations for $n(\bi{r})$ in terms of $V(\bi{r})$ enter on
the right-hand side in \Eq{2a}.
The partner equation (\ref{eq:2b}) requires a corresponding approximation for
the IEF,
\begin{equation}
\label{eq:46}
E_{\mathrm{int}}[n]=\half\int(\D\bi{r})(\D\bi{r}')\,
V_{\mathrm{pair}}{\left(\boldsymbol{|}\bi{r}-\bi{r}'\boldsymbol{|}\right)}
n^{(2)}(\bi{r},\bi{r}'; \bi{r},\bi{r}')\,,
\end{equation}
where $V_{\mathrm{pair}}(a)$ is the potential energy of an interacting pair of
particles separated by distance $a$.
For simplicity we are here content with considering spin-independent forces
between the particles;
spin-dependent forces, such as those between dipolar atoms, can be dealt with
as well \cite{Goral+2:01,Fang+1:11}.
We can use any available approximate IEF for this purpose or, better, search
for an IEF in an approximation that is consistent with that for the KEF (or,
rather, its functional derivative).
We are thus assigned the task of expressing the two-particle density
$n^{(2)}(\bi{r},\bi{r}'; \bi{r},\bi{r}')$ in terms of the single-particle
density $n(\bi{r})=n^{(1)}(\bi{r};\bi{r})$.

Although this matter is beyond the scope of this article, let us briefly
mention a possible strategy. 
Following Dirac \cite{Dirac:30}, we accept the approximation
\begin{equation}
\label{eq:47}
n^{(2)}(\bi{r},\bi{r}'; \bi{r},\bi{r}')\simeq n(\bi{r})n(\bi{r'})
-\half n^{(1)}(\bi{r};\bi{r}')n^{(1)}(\bi{r}';\bi{r})\,,
\end{equation}
which corresponds to splitting the IEF into a direct energy and an exchange
energy contribution.
For the single-particle density matrix $n^{(1)}(\bi{r};\bi{r}')$ we 
employ what \Eq{5} suggests, that is $n^{(1)}(\bi{r};\bi{r}')\simeq%
2\bra{\bi{r}}\eta{\bigl(\mu-H(\bi{P},\bi{R})\bigr)}\ket{\bi{r}'}$,
then relate the step function of $\mu-H$ to the evolution operator, 
to which we apply a suitable ST factorization.
This yields $n^{(1)}(\bi{r};\bi{r}')$ in terms of $V(\bi{r})$, and to
complete the job we have to invert the mapping 
$V(\bi{r})\mapsto n(\bi{r})$ in \Eq{2a} in a consistent approximation.
For the TF approximation, for example, this results in
\begin{eqnArray}
\label{eq:48}
\ds\frac{\delta E_{\mathrm{int}}[n]}{\delta n(\bi{r})}&=&\ds
\int(\D\bi{r}')\,V_{\mathrm{pair}}(r')n(\bi{r}+\bi{r}')
\\[2ex]&&\ds\mbox{}
-\int\limits_0^{\infty}\frac{\D r'}{r'}V_{\mathrm{pair}}(r')\,
z\,\J_{\half D}(z)\,\J_{\half D-1}(z)\Bigr|_{z=r'Q(\bi{r})}
\end{eqnArray}
with $Q(\bi{r})=\sqrt{4\pi}{\left[\half{\left(\half D\right)}!\,%
n(\bi{r})\right]}^{\frac{1}{D}}$, for use in \Eq{2b};
the first integral is the Hartree term of \Eq{EintH}, there stated for the
Coulomb interaction of the Hooke atom.
We shall return to these matters in due course.

\section{Digression: Beyond the leapfrog approximation}
\begin{figure}
\begin{center}
\includegraphics[width=0.7\linewidth]{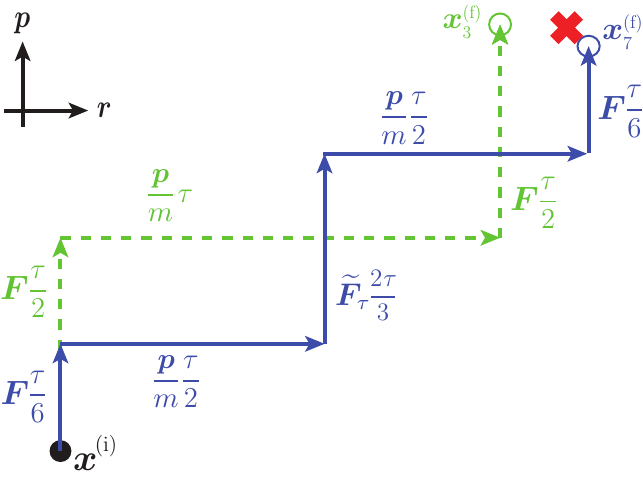}
\caption{\label{LeapFrog}%
   Illustration of the finite-difference schemes that correspond to $U_3$ (green
   dashed line, standard second-order leapfrog) and $U_7$ (blue solid
   line, new fourth-order leapfrog), respectively. 
   They propagate the initial (i) to the approximate final (f) phase-space
   points within time $\tau$. The red cross marks the exact final phase-space
   point.}   
\end{center}
\end{figure}
Besides supplying approximate solutions to the time-dependent Schr\"{o}dinger
equation in  quantum mechanics, the split-operator approximations for the
unitary evolution operator such as $U_3$ and $U_7\bigr|_{\epsilon\to0}$ have
counter parts in the context of symplectic approximations in solving
Hamilton's equations of motion in classical mechanics.
One obtains an approximate solution for
\begin{equation}
\label{eq:LF1}
\Dt\bi{r}(t)=\frac{1}{m}\bi{p}(t)\,,\ \quad
\Dt\bi{p}(t)=\bi{F}\bigl(\bi{r}(t)\bigr)=-\grad V\bigl(\bi{r}(t)\bigr)
\end{equation}
by the so-called leapfrog algorithm \cite{Newton:87} 
(see also, e.g., \cite{Neal:11}):
\begin{equation}
\label{eq:LF2}
\begin{tabular}[b]{rl}
(i)   & \mbox{use $\ds\Dt\bi{r}=0$, $\ds\Dt\bi{p}=\bi{F}(\bi{r})$
	for duration $\half\tau$;}\\[1.5ex]
(ii)  & \mbox{use $\ds\Dt\bi{r}=\frac{1}{m}\bi{p}$, $\ds\Dt\bi{p}=0$
	for duration $\tau$;}\\[1ex]
(iii) & repeat step (i);
\end{tabular}
\end{equation}
which takes one from time $t$ to time $t+\tau$ (see the green dashed lines in
fig\-ure~\ref{LeapFrog}). 
Clearly, this second-order leapfrog is the analog of the ST3 factorization in
\Eq{9}. 
There is also a leapfrog analog of ST3'.

The fourth-order approximation in \Eq{U7inf} yields a corresponding fourth-order
leapfrog (see the blue solid lines in figure~\ref{LeapFrog}):
\begin{equation}
\label{eq:LF3}
\begin{tabular}[b]{rl}
(i)  & \mbox{use $\ds\Dt\bi{r}=0$, $\ds\Dt\bi{p}=\bi{F}(\bi{r})$
	for duration $\frac{1}{6}\tau$;}\\[1.5ex]
(ii) & \mbox{use $\ds\Dt\bi{r}=\frac{1}{m}\bi{p}$, $\ds\Dt\bi{p}=0$
	for duration $\half\tau$;}\\[1.5ex]
(iii) & \mbox{use $\ds\Dt\bi{r}=0$, 
	$\ds\Dt\bi{p}=\widetilde{\!\bi{F}}_{\hspace{-0.15em}\tau}(\bi{r})$
	for duration $\frac{2}{3}\tau$;}\\[1.5ex]
(iv) & repeat step (ii);\\
(v)  & repeat step (i);
\end{tabular}
\end{equation}
where
\begin{equation}
\label{eq:LF4}\fl\rule{5em}{0pt}
\widetilde{\!\bi{F}}_{\hspace{-0.15em}\tau}(\bi{r})
=-\grad{\left(V-\frac{1}{48m}{\left[\tau\grad V\right]}^2\right)}(\bi{r})
={\left(\bi{F}+\frac{\tau^2}{24m}\bi{F}\cdot\grad\bi{F}\right)}(\bi{r})
\end{equation}
in step (iii). 
This fourth-order leapfrog promises substantial improvements over the
second-order leapfrog. 
We are currently exploring this territory with very encouraging initial
results \cite{EgeUROPS}.

\section{Summary and outlook}
We presented a method for incorporating inhomogeneity corrections into the
approximate single-particle density without resorting to a gradient expansion.
By insisting on no error for a constant force, we ensure that Langer's
correction is properly accounted for.
This improves the density very much at the transition from the classically
allowed to the classically forbidden region, and beyond, as is confirmed in the
examples of our benchmarking exercise.

In order to advance the agenda, we shall next develop corresponding
approximations for the interaction-en\-er\-gy functional, thereby setting the
stage for the computation of self-consistent approximate solutions for a
variety of fermion systems, among them the electron gas in atoms, molecules,
and solids, as well as ultracold neutral atoms in optical traps. 
Generally, we expect the computational effort for our orbital-free approach to
be linear in system size --- in contrast to the cubic scaling of Kohn-Sham
calculations in the self-consistent scheme. 
Some of the applications are likely to require the use of pseudopotentials
\cite{Pickett:89,Legrain+1:15}.
Further, it will be interesting to investigate the analogous approximations
for momentum-space densities 
\cite{Henderson:81,Englert:92,Cinal+1:92,Cinal+1:93}, which are directly
accessible in experiments with degenerate atom gases.

\section*{Acknowledgments}
JHH acknowledges the financial support of the Graduate School for Integrative
Sciences \& Engineering at the National University of Singapore. 
This work is funded by the Singapore Ministry of Education and the National
Research Foundation of Singapore.

\appendix

\section{Wigner function}\manuallabel{Wigner}{A}
The fourth-order approximation in \Eq{U7inf} gives rise to an approximate Wigner
function of the time-evolution operator 
\begin{eqnArray}[4em]\label{eq:A1}
&&\ds{\left(\exp{\left(-\frac{\I t}{\hbar}
	{\left[\frac{1}{2m}\bi{P}^2+V(\bi{R})\right]}\right)}
\right)}_{\mathrm{W}}(\bi{r},\bi{p})\\[3ex]
&\cong&\ds{\left(\frac{m}{2\pi\hbar\I t}\right)}^D
\int(\D \bi{s})(\D\bi{a})\,
\exp{\left(\frac{\I m}{2\hbar t}{\left(\bi{a}^2+\bi{s}^2\right)}\right)}
\,\exp{\left(\frac{\I}{\hbar}\bi{s}\cdot\bi{p}\right)}\\[3ex]
&&\ds\times
\exp{\left(-\frac{\I t}{\hbar}{\left[\frac{1}{6}V(\bi{r}+\frac{1}{2}\bi{s})
	+\frac{1}{6}V(\bi{r}-\frac{1}{2}\bi{s})
	+\frac{2}{3}\widetilde{V_t}(\bi{r}+\frac{1}{2}\bi{a})\right]}\right)}
\end{eqnArray}
with
\begin{equation}\label{eq:A2}
\widetilde{V_t}(\bi{r})=V(\bi{r})
-\frac{1}{48m}{\bigl[t\grad V(\bi{r})\bigr]}^2\,.
\end{equation}
One reverts to the leading-order gradient correction of the Wigner function by
expanding $V(\bi{r}\pm\frac{1}{2}\bi{s})$ and
$\widetilde{V_t}(\bi{r}+\frac{1}{2}\bi{a})$ around $\bi{r}$ up to second
order in $\grad$ and then evaluating the resulting Gaussian integrals. 
Keeping only terms through second order in $\grad$, we so arrive at
\begin{eqnArray}\label{eq:A3}
&&\ds{\left(\exp{\left(-\frac{\I t}{\hbar}
		{\left[\frac{1}{2m}\bi{P}^2+V(\bi{R})\right]}\right)}
	\right)}_{\mathrm{W}}(\bi{r},\bi{p})\\[3ex]
&\cong&\ds{\left[1+\frac{t^2}{8m}\grad^2V(\bi{r})\right]}
\exp{\left(-\frac{\I t}{\hbar}{\left[\frac{\bi{p}^2}{2m}
     +V(\bi{r})\right]}\right)}\\[3ex]
&&\ds\times\exp{\left(-\frac{\I t}{\hbar}\frac{1}{24}
	{\left({\left[\frac{t}{m}\bi{p}\cdot\grad\right]}^2V(\bi{r})
	+\frac{1}{m}{\bigl[t\grad V(\bi{r})\bigr]}^2\right)}\right)}\,.
\end{eqnArray}
This agrees with the approximate Wigner function of $U(t)$ in 
appendix~B of \cite{Trappe+3:17}, there obtained from a gradient
expansion.  
We now regard this as an approximation to \Eq{A1}, which improves on the
gradient expansion.

\section{Derivation of the approximate densities}\manuallabel{nDerivations}{B}
One way of deriving $n_7$ proceeds from the combination of 
\Eq[equations~]{25} and \Eq[]{7}, i.e.,
\begin{eqnArray}[0.3em]\label{eq:B1}
 n_7(\bi{r})&=&\makebox[0pt][l]{$
\ds 2\Int\frac{\D t}{2\pi\I t}
\int\!\frac{(\D \bi{r}_1)(\D \bi{r}_2)}{(2\pi\hbar)^{3D}}\,
\exp{\left(\frac{\I t}{\hbar}{\bigr[\mu-x_1V(\bi{r}_1)-x_2V(\bi{r}_2)
  -(x_3+x_4)V(\bi{r})\bigr]}\right)}$}\\[3ex]
&&\ds\hphantom{2\int\D}
\times\int\!(\D \bi{p}_1)(\D\bi{p}_0)(\D\bi{p}_2)
\,\exp{\left(-\frac{\I t}{2\hbar m}
{\left(y_1\bi{p}_1^2+y_0\bi{p}_0^2+y_2\bi{p}_2^2\right)}\right)}
\\[3ex]&&\ds\hphantom{ 2\Int\frac{\D t}{2\pi\I t}}\times 
\exp{\left(\frac{\I}{\hbar}{\bigl[\bi{p}_1\cdot(\bi{r}-\bi{r}_1)
  +\bi{p}_0\cdot(\bi{r}_1-\bi{r}_2)+\bi{p}_2\cdot(\bi{r}_2
  -\bi{r})\bigr]}\right)}\,,
\end{eqnArray}
where the evaluation of the Gaussian $\bi{p}$ integrals results in
\begin{eqnArray}[2em]\label{eq:B2}
 n_7(\bi{r})&=&\ds 2\Int\frac{\D t}{2\pi\I t}
\int\!\frac{(\D \bi{r}_1)(\D \bi{r}_2)}{(2\pi\hbar)^{3D}}
   {\left(\frac{2\pi\hbar}{\I t}\right)}^{\frac{3}{2}D}
   {\left(\frac{m^3}{y_0y_1y_2}\right)}^{\frac{1}{2}D}
   \exp{\left(\frac{\I}{\hbar}\frac{m}{2t}a^2\right)}\,
\\[3ex]&&\ds\hphantom{ 2\Int\frac{\D t}{2\pi\I t}}
\times\exp{\left(\frac{\I t}{\hbar}{\bigl[\mu-x_1V(\bi{r}_1)-x_2V(\bi{r}_2)
	-(x_3+x_4)V(\bi{r})\bigr]}\right)}
\end{eqnArray}
with $a^2$ as in \Eq{39}.
We use the identity
\begin{equation}\label{eq:B3}
\fl\rule{2em}{0pt} {\left(\frac{a m}{\I t}\right)}^{\frac{3}{2}D}
\exp{\left(\frac{\I}{\hbar}\frac{m}{2t}a^2\right)}=
\frac{a}{\hbar}\int\limits_{0}^{\infty}\!\D p\,p^{\frac{3}{2}D}
\exp{\left(-\frac{\I t}{\hbar}\frac{p^2}{2m}\right)} 
\J_{\frac{3D}{2}-1}{\left(\frac{ap}{\hbar}\right)}
\end{equation}
to make the exponent linear in $t$, after which the resulting $t$ integral
yields a step function and \Eq{B2} becomes
\begin{equation}\label{eq:B4}
\fl\rule{3em}{0pt} 
n_7(\bi{r})=2\int\!\frac{(\D\bi{r}_1)(\D\bi{r}_2)}{(y_0y_1y_2)^{\frac{1}{2}D}}\,
\frac{a}{\hbar}\int\limits_{0}^{\infty}\D p\,
{\left(\frac{p}{2\pi\hbar a}\right)}^{\frac{3}{2}D}
\J_{\frac{3D}{2}-1}{\left(\frac{ap}{\hbar}\right)}\eta{\left(p_{7}^2-p^2\right)}
\end{equation}
where
\begin{equation}\label{eq:B5}
p_{7}^2=2m{\bigl[\mu-x_1V(\bi{r}_1)-x_2V(\bi{r}_2)-(x_3+x_4)V(\bi{r})\bigr]}\,.
\end{equation}
The elementary $p$ integration now establishes \Eq{38} with $k_5$ replaced by
$k_7$ as in \Eq{41}. 

Quite similarly, we get the analog of \Eq{B2} for the density
corresponding to the fourth-order approximation $U_7\bigr|_{\epsilon\to0}\,$
in \Eq{U7inf}, 
\begin{eqnArray}\label{eq:B6}
n_7(\bi{r})\Bigr|_{\epsilon\to0}&=&
\ds 2\int(\D\bi{a})\!\Int\frac{\D t}{(2\pi\I t)^{D+1}}
{\left(\frac{2m}{\hbar}\right)}^D
\exp{\left(\frac{\I}{\hbar}\frac{2m}{t}\bi{a}^2\right)}\\[3ex]
&&\ds\hphantom{2\int(\D\bi{a})\!\Int}
\times\exp{\left(\frac{\I t}{\hbar}{\left[\mu -\frac{1}{3}V(\bi{r})
    -\frac{2}{3}\widetilde{V}_t(\bi{r}+\bi{a})\right]}\right)}
\end{eqnArray}
with $\widetilde{V}_t(\bi{r})$ of \Eq{A2}.
We substitute
\begin{equation}\label{eq:B7}
\frac{2m}{\hbar t}\bi{a}^2=\frac{2ma^2}{\hbar t}=\frac{1}{4s} 
\qquad{\mathrm{or}}\qquad t=\frac{8ma^2}{\hbar}s
\end{equation}
and arrive at
\begin{equation}\label{eq:B8}
\fl\rule{2em}{0pt} n_7^{\ }(\bi{r})\Bigr|_{\epsilon\to0}=
2\int\!(\D\bi{a})\,\frac{1}{(4\pi a^2)^D}\,\frac{1}{\pi}
\Int\frac{\D s}{(2\I s)^{D+1}}\,
\exp{\left(\I As+\frac{\I}{3}(Bs)^3+\frac{\I}{4s}\right)}
\end{equation}
with
\begin{equation}\label{eq:B9}
A=A(\bi{r},\bi{a})
={\left(\frac{2a}{\hbar}\right)}^2{\Bigl[2m{\left(\mu-\case{1}{3}V(\bi{r})
-\case{2}{3}V(\bi{r}+\bi{a})\right)}\Bigr]}
\end{equation}
and
\begin{equation}\label{eq:B10}
B=B(\bi{r},\bi{a})={\left(\frac{2a}{\hbar}\right)}^2
{\left(\case{1}{3}{\bigl[\hbar m\grad V(\bi{r}+\bi{a})\bigr]}^2
\right)}^{1/3}\,.
\end{equation}

It is useful to introduce the bivariate special function $K_D(A,B)$ in
accordance with
\begin{equation}\label{eq:B11}
K_D(A,B)=\frac{1}{\pi}\Int\frac{\D s}{(2\I s)^{D+1}}\,\Exp{\I \varphi(s)}
\end{equation}
for 
\begin{equation}\label{eq:B12}
\varphi(s)=As+\frac{1}{3}(Bs)^3+\frac{1}{4s}\,;
\end{equation}
then
\begin{equation}\label{eq:B13}
n_7^{\ }(\bi{r})\Bigr|_{\epsilon\to0}=
2\int\frac{(\D\bi{a})}{(4\pi a^2)^D}K_D(A,B)\,.
\end{equation}
After rewriting \Eq{B11} with the help of Airy averaging,
\begin{equation}\label{eq:B14}
\fl\rule{3em}{0pt} 
K_D(A,B)=\int_{-\infty}^{\infty}\D x\,\Ai(x)\;
\frac{1}{\pi}\Int\frac{\D s}{(2\I s)^{D+1}}\,
\exp{\left(\I (A-xB)s+\frac{\I}{4s}\right)}\,,
\end{equation}
we can close the contour of the $s$ integration by a large-radius semicircle
in the lower half-plane when $A-xB<0$, and by a semicircle in the upper
half-plane when $A-xB>0$. 
Since the only singularity of the integrand is at $s=0$, we get a null result
for $K_D(A,B)$ if $A-xB<0$ and pick up the residue if $A-xB>0$,
\begin{eqnArray}\label{eq:B15}
 K_D(A,B)&=&\ds\int_{-\infty}^{\infty}\D x\,\Ai(x)\,\eta(A-xB)
\\[3ex]
&&\ds\quad\ \ \times{\left[\mathrm{Res}_{s=0}{\left(\frac{2\I}{(2\I s)^{D+1}}
\exp{\left(\I (A-xB)s+\frac{\I}{4s}\right)}\right)}\right]}\,.
\end{eqnArray}
A generating function of the Bessel functions,
\begin{equation}\label{eq:B16}
\fl\rule{3em}{0pt} 
\exp{\left(\half z{\left(u-u^{-1}\right)}\right)}=\sum_{k=-\infty}^{\infty}u^k\J_k(z)
=\sum_{k=-\infty}^{\infty}u^{D+k}\J_{D+k}(z)\,,
\end{equation}
identifies the residue in
\begin{eqnArray}\label{eq:B17}
&&\ds\frac{2\I}{(2\I s)^{D+1}}\,\exp{\left(\I(A-xB)s+\frac{\I}{4s}\right)}
\\[3ex]&=&\ds\frac{1}{s}\sqrt{A-xB}^D
\sum_{k=-\infty}^{\infty}{\left[2\I\sqrt{A-xB}s\right]}^k
\J_{D+k}{\left(\sqrt{A-xB}\,\right)}
\end{eqnArray}
as the $k=0$ term.
It follows that
\begin{eqnArray}\label{eq:B18}
K_D(A,B)&=&\ds\int_{-\infty}^{\infty}\D x\,\Ai(x)\,
\eta{\left(A-xB\right)}\sqrt{A-xB}^D \J_D{\left(\sqrt{A-xB}\,\right)}\\[3ex]
&=&\ds\int_{-\infty}^{\infty}\D x\,\Ai(x)\,
{\left(2ak^{(x)}_{7}\right)}^D\J_D{\left(2ak^{(x)}_{7}\right)}
\end{eqnArray}
with $k^{(x)}_{7}$ as in \Eq{43}. 
We obtain \Eq{42} when using the last expression for $K_D(A,B)$ in \Eq{B13}.

\section{How to compute $K_D(A,B)$ numerically}\manuallabel{KDnumerics}{C}
The integral representations for $K_D(A,B)$ in \Eq[equations~]{B11} and
\Eq[]{B18} are not well suited for a numerical evaluation.
To establish a numerically tractable form we proceed from \Eq{B11} and note
that $A$ is real and $B$ positive in \Eq{B12} and
\begin{equation}\label{eq:C1}
\varphi'(s)=\frac{\D\varphi(s)}{\D s}=A+B^3s^2-\frac{1}{4s^2}\,.
\end{equation}
For $s>0$, we have ${\varphi(s)\geq\varphi_0=\varphi(s_0)}$,
$\varphi'(s_0)=0$ with
\begin{equation}\label{eq:C2}
s_0^2=\frac{1}{2B^3}{\left(\sqrt{A^2+B^3}-A\right)}
=\frac{1}{2}\frac{1}{\sqrt{A^2+B^3}+A}\,.
\end{equation}
The two positive solutions of ${\varphi(s)=\phi>\varphi_0}$ are denoted
$s_1(\phi)$ and $s_2(\phi)$, respectively, with ${0<s_1(\phi)<s_0<s_2(\phi)}$
and ${s_1'(\phi)<0<s_2'(\phi)}$.
For symmetry reasons, we have
\begin{eqnArray}[4em]\label{eq:C3}
K_D(A,B)&=&\ds\frac{2}{\pi}\,\Real\int_{-\I\epsilon}^{\infty}
\frac{\D s}{(2\I s)^{D+1}}\,\Exp{\I\varphi(s)}\\[3ex]
&=&\ds
\frac{1}{\pi D}\,\Real\int_{-\I\epsilon}^{\infty}\D s\;
\exp{\left(\I{\left[\varphi(s)-D\frac{\pi}{2}\right]}\right)}
\pder{}{s}\frac{\I}{(2s)^D}\,,
\end{eqnArray}
if the integration path crosses the imaginary axis at ${s=-\I\epsilon}$ with
${\epsilon>0}$. 
We note that $\I\varphi(-\I\epsilon)$ is real and also that
${s^n\Exp{\I\varphi(s)}\bigr|_{s=-\I\epsilon}\to0}$ as ${0<\epsilon\to0}$ for all
powers $n$, negative or positive. It is expedient to integrate from
$s=-\I\epsilon$ to $s=s_0$ and then along 
the real axis from $s_0$ to $\infty$.
This decomposes $K_D(A,B)$ into two pieces, 
\begin{equation}\label{eq:C4}
K_D(A,B)=K_D^{(1)}(A,B)+K_D^{(2)}(A,B)\,,
\end{equation}
the contributions respectively associated with ${s_1(\phi)}$ and ${s_2(\phi)}$: 
\begin{eqnArray}[5em]\label{eq:C5}
K_D^{(1)}(A,B)&=&\ds\frac{1}{\pi D}\,\Real\int_{-\I\epsilon}^{s_0}\D s\,
\exp{\left(\I{\left[\varphi(s)-D\frac{\pi}{2}\right]}\right)}
\pder{}{s}\frac{\I}{(2s)^D}\,,\\[3ex] 
K_D^{(2)}(A,B)&=&\ds\frac{1}{\pi D}\Real\int_{s_0}^{\infty}\D s\,
\exp{\left(\I{\left[\varphi(s)-D\frac{\pi}{2}\right]}\right)}
\pder{}{s}\frac{\I}{(2s)^D}\,.
\end{eqnArray}
For the integration by parts in the next step, we note the identity
\begin{eqnArray}[4em]\label{eq:C7}
&&\ds \exp{\left(\I{\left[\varphi(s)-D\frac{\pi}{2}\right]}\right)}
\pder{}{s}\frac{1}{(2s)^D}\\[3ex]
&=&\ds
\pder{}{s}{\left[\exp{\left(\I{\left[\varphi(s)-D\frac{\pi}{2}\right]}\right)}
\left(\frac{1}{(2s)^D}-g\bigl(\varphi(s)\bigr)\right)\right]}\\[3ex]
&&\ds\mbox{}-\I\varphi'(s)\,
\exp{\left(\I{\left[\varphi(s)-D\frac{\pi}{2}\right]}\right)}
{\left(\frac{1}{(2s)^D}-f\bigl(\varphi(s)\bigr)\right)}
\end{eqnArray}
with any $g(\phi)$ of our liking and $f(\phi)=g(\phi)-\I g'(\phi)$.
For $K_D^{(2)}(A,B)$, where $s$ is real throughout and the
essential singularity at $s=0$ plays no role, we choose $g(\phi)=0$ and obtain
\begin{eqnarray}\label{eq:C8}
\fl\rule{0.2em}{0pt} K_D^{(2)}(A,B)= \frac{1}{\pi D}\,\Real{\left(
-\frac{\I\exp{\left(\I{\left[\varphi_0-D\frac{\pi}{2}\right]}\right)}}{(2s_0)^D}
+\int_{\varphi_0}^{\infty}\D\phi\,
\frac{1}{[2s_2(\phi)]^D}\,
\exp{\left(\I{\left[\phi-D\frac{\pi}{2}\right]}\right)}\right)}\nonumber\\
\end{eqnarray}
after switching from integration over $s$ to integration over $\phi$.
For $K_D^{(1)}(A,B)$, where we need to watch out for the singularity at $s=0$,  
we choose a real polynomial in $\phi$ for $f(\phi)$ such that
\begin{eqnArray}[5em]\label{eq:C9}
&&\ds\frac{1}{(2s)^D}-f{\bigl(\varphi(s)\bigr)}\to0\quad 
\mbox{as}\quad s\to0\\[3ex]\mbox{or}&&\ds
\frac{1}{[2s_1(\phi)]^D}-f(\phi)\to0\quad 
\mbox{as}\quad \phi\to\infty\,,\\
\end{eqnArray}
and 
\begin{equation}\label{eq:C10}
g(\phi)=\I\int_0^\phi \D\phi'\,\exp\bigl(-\I[\phi-\phi']\bigr)f(\phi')\,.
\end{equation}
Then, the $\epsilon\to0$ limit in $K_D^{(1)}(A,B)$ is well defined, and we
obtain
\begin{eqnArray}[1ex]\label{eq:C11}
 K_D^{(1)}(A,B)&=&\ds\frac{1}{\pi D}\,\Real\Biggl(
\frac{\I\exp{\left(\I{\left[\varphi_0-D\frac{\pi}{2}\right]}\right)}}{(2s_0)^D}
-\I\exp{\left(\I{\left[\varphi_0-D\frac{\pi}{2}\right]}\right)}
g(\varphi_0)\\[3ex]&&\ds\hphantom{\frac{1}{\pi D}\,\Real\Biggl(}
-\int_{\varphi_0}^{\infty}\D\phi\,
{\left(\frac{1}{[2s_1(\phi)]^D}-f(\phi)\right)}
\,\exp{\left(\I{\left[\phi-D\frac{\pi}{2}\right]}\right)}\Biggr)\\
&=&\ds\frac{1}{\pi D}\,\Real\Biggl(
\frac{\I\exp{\left(\I{\left[\varphi_0-D\frac{\pi}{2}\right]}\right)}}{(2s_0)^D}
+\int_0^{\varphi_0} \D\phi\;\exp{\left(\I{\left[\phi-D\frac{\pi}{2}\right]}
\right)}f(\phi)\\[3ex]
&&\ds\hphantom{\frac{1}{\pi D}\,\Real\Biggl(}
-\int_{\varphi_0}^{\infty}\D\phi\,
{\left(\frac{1}{[2s_1(\phi)]^D}-f(\phi)\right)}
\,\exp{\left(\I{\left[\phi-D\frac{\pi}{2}\right]}\right)}\Biggr)\,.
\end{eqnArray}
Together with \Eq{C8}, this yields
\begin{eqnArray}[1em]\label{eq:C12}
K_D(A,B)&=&\ds\frac{1}{\pi D}\int_{-\infty}^{\infty} \D\phi\,
\cos{\left(\phi-D\frac{\pi}{2}\right)}\\[3ex]
&&\ds\hphantom{\frac{1}{\pi D}\int}
\times{\left[\eta(\phi)f(\phi)-\eta(\phi-\varphi_0)
{\left(\frac{1}{[2s_1(\phi)]^D}-\frac{1}{[2s_2(\phi)]^D}\right)}\right]}\,,
\end{eqnArray}
where $f(\phi)$ has to be chosen in accordance with \Eq{C9}.
In view of the $\phi$-expansion of
\begin{eqnArray}[5em]\label{eq:C13}
\ds\frac{1}{2s_1}&=&\ds 2\phi-2As_1-\frac{2}{3}{\left(Bs_1\right)}^3
=2\phi-A\,(2s_1)-\frac{1}{12}\bigl[B\,(2s_1)\bigr]^3\\[3ex]
&=&\ds
2\phi-\frac{A}{2\phi}-\frac{A^3+B^3/12}{(2\phi)^3}+\mathcal{O}(\phi^{-5})
\end{eqnArray}
for large positive $\phi$,
we take
\begin{equation}\label{eq:C14}
f(\phi)=\cases{2\phi&for $D=1$\,,\\
(2\phi)^2-2A &for $D=2$\,,\\
(2\phi)^3-6A\phi &for $D=3$\,.\\}
\end{equation}
It follows that
\begin{equation}\label{eq:C15}
\fl\rule{2em}{0pt} 
K_1(A,B)=\frac{2}{\pi}\int_{-\infty}^{\infty}\D\phi\,
{\left[\phi\,\eta(\phi)-{\left(\frac{1}{4s_1(\phi)}-\frac{1}{4s_2(\phi)}\right)}
\eta(\phi-\varphi_0)\right]}\sin\phi\,, 
\end{equation}
and
\begin{eqnArray}[2em]\label{eq:C16}
K_2(A,B)&=&\ds\frac{2}{\pi}\int_{-\infty}^{\infty}\D\phi\,
\Biggl[{\left(-\phi^2+\half A\right)}\eta(\phi) \\[3ex]&&\ds
\hphantom{\frac{2}{\pi}\int_{-\infty}^{\infty}\D\phi\,\Biggl]}
+{\left(\frac{1}{[4s_1(\phi)]^2}-\frac{1}{[4s_2(\phi)]^2}\right)}
\eta(\phi-\varphi_0)\Biggr]\cos\phi\,,
\end{eqnArray}
as well as
\begin{eqnArray}[2em]\label{eq:C17} 
K_3(A,B) 
&=&\ds\frac{8}{3\pi}\int_{-\infty}^{\infty}\D\phi\,
\Biggl[{\left(-\phi^3+\frac{3}{4}A\phi\right)}\eta(\phi)\\[3ex]&&\ds
\hphantom{\frac{8}{3\pi}\int_{-\infty}^{\infty}\D\phi\,\Biggl]}
+{\left(\frac{1}{[4s_1(\phi)]^3}-\frac{1}{[4s_2(\phi)]^3}\right)}
\eta(\phi-\varphi_0)\Biggr]\sin\phi\,.
\end{eqnArray}
\Eq[Equations~]{C15}--\Eq[]{C17} are well suited for a numerical evaluation.

\end{document}